\documentclass[aps,prb,twocolumn,floatfix,longbibliography,superscriptaddress]{revtex4-2}

\usepackage{amsmath}
\usepackage{amssymb}
\usepackage{times}
\usepackage{braket}
\usepackage[pdftex]{graphicx}
\usepackage{color}
\usepackage{blindtext}
\usepackage{nicefrac}
\usepackage[colorlinks=true, citecolor=blue, urlcolor=blue, linkcolor=blue]{hyperref}

\usepackage{todonotes}

\renewcommand{\vec}[1]{\boldsymbol{#1}}

\newcommand{\change}[1]{\textcolor{black}{#1}}
\newcommand{\changeb}[1]{\textcolor{black}{#1}}

\renewcommand{\ket}[1]{\lvert#1\rangle} 
\newcommand{\braopket}[3]{\langle #1 | #2 | #3\rangle} 

\begin{document}

\title{Chirality-induced orbital Edelstein effect in an analytically solvable model}

\author{B{\"o}rge G{\"o}bel}
\email[Correspondence email address: ]{boerge.goebel@physik.uni-halle.de}
\affiliation{Institut f\"ur Physik, Martin-Luther-Universit\"at Halle-Wittenberg, D-06099 Halle (Saale), Germany}

\author{Lennart Schimpf}
\affiliation{Institut f\"ur Physik, Martin-Luther-Universit\"at Halle-Wittenberg, D-06099 Halle (Saale), Germany}

\author{Ingrid Mertig}
\affiliation{Institut f\"ur Physik, Martin-Luther-Universit\"at Halle-Wittenberg, D-06099 Halle (Saale), Germany}

\date{\today}

\begin{abstract}
Chirality-induced spin selectivity (CISS), a phenomenon wherein chiral structures selectively determine the spin polarization of electron currents flowing through the material, has garnered significant attention due to its potential applications in areas such as spintronics, enantioseparation, and catalysis. The underlying physical effect is the Edelstein effect that converts charge to angular momentum. Besides a spin contribution there exists a contribution based on the orbital angular momentum but the precise mechanism for its generation remains yet to be understood. Here, we introduce the minimal model for explaining the phenomenon based on the orbital Edelstein effect. We consider non-local inter-site contributions to the current-induced orbital angular momentum and reveal the underlying mechanism by analytically calculating the Edelstein susceptibilities in a tight-binding and Boltzmann approach. While the orbital angular momentum is directly generated by the chirality of the crystal, the spin contribution of each spin-split band pair relies on spin-orbit coupling. Using tellurium as an example, we show that the orbital contribution surpasses the spin contribution by orders of magnitude.
\end{abstract}

\maketitle


\section{Introduction} \label{sec:introduction}

\begin{figure}[t!]
    \centering
    \includegraphics[width=1\columnwidth]{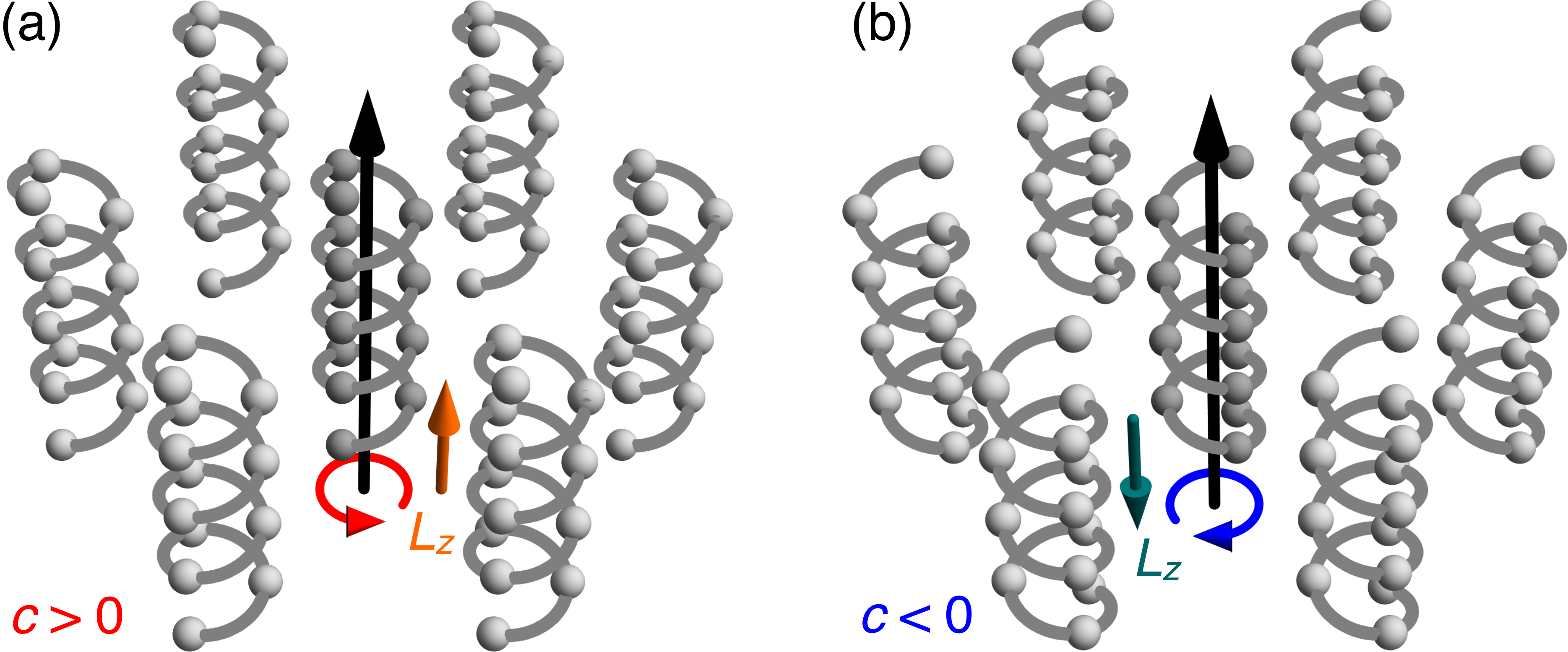}
    \caption{Chiral helices. (a) Right-handed helix structure characterized by $c>0$. (b) Left-handed structure characterized by $c<0$. The black arrow symbolizes the translational motion of electrons along $z$. It is constrained to a circular motion (red and blue) which generates opposite orbital angular momentum $L_z$ \change{(orange and green)} for the two opposite crystal chiralities.}
    \label{fig:overview}
\end{figure}

Chirality is a geometric property that affects a wide range of scientific fields, from chemistry and biology to physics and material engineering. A structure is chiral if it is distinguishable from its mirror image; it lacks mirror, inversion and roto-inversion symmetries~\cite{furukawa2021current}. 
The structural chirality of molecules like the DNA molecule or crystals like tellurium and selenium [cf. Fig.~\ref{fig:overview}] can affect biochemical properties and even couples to physical observables like the angular momentum which gives rise to \change{intriguing phenomena like electrical magnetochiral anisotropy~\cite{rikken2001electrical,pop2014electrical,aoki2019anomalous,inui2020chirality} and} chirality-induced spin selectivity (CISS)~\cite{ray1999asymmetric,lu2019spin,huang2020magneto,furukawa2021current,calavalle2022gate,inui2020chirality,nabei2020current,dalum2019theory,geyer2020effective,cheong2022magnetic,naskar2023chiral}. The chirality of a structure determines the sign of the spin-polarization of electrons moving through it. CISS was first observed in chiral molecules~\cite{ray1999asymmetric}. More recently, this phenomenon has been observed in a broader range of systems, including hybrid organic-inorganic perovskites~\cite{lu2019spin,huang2020magneto} and chiral crystals like tellurium~\cite{furukawa2017observation,furukawa2021current,calavalle2022gate} or CrNb$_3$S$_6$~\cite{inui2020chirality,nabei2020current}. 

Despite extensive experimental observations, the underlying physical mechanisms governing CISS remain a topic of active debate. \change{Some} theoretical explanations are based on the charge-to-spin conversion effect named inverse spin-galvanic effect or Edelstein effect (EE)~\cite{aronov1989nuclear,edelstein1990spin}: Due to a broken inversion symmetry, the spin polarizations of electrons with opposite wave vectors differ which leads to the generation of a spin density once an electric field or currents are applied. The EE typically occurs at interfaces where the broken inversion symmetry gives rise to a Rashba-type spin-orbit coupling~\cite{rashba1960properties,bychkov1984properties,bychkov1984oscillatory}. In a chiral material, however, the inversion symmetry is innately broken and a collinear effect can occur where the spins are parallel to the wave vector~\cite{furukawa2017observation,furukawa2021current,calavalle2022gate,roy2022long,slawinska2023spin} and the theoretically calculated Edelstein susceptibilities indeed depend on the chirality of the structure. However, as was pointed out in Refs.~\cite{roy2022long,slawinska2023spin}, they are an order of magnitude smaller than what was measured experimentally in tellurium~\cite{furukawa2017observation} pointing towards the existence of important additional contributions. 

\begin{figure}[t!]
    \centering
    \includegraphics[width=1\columnwidth]{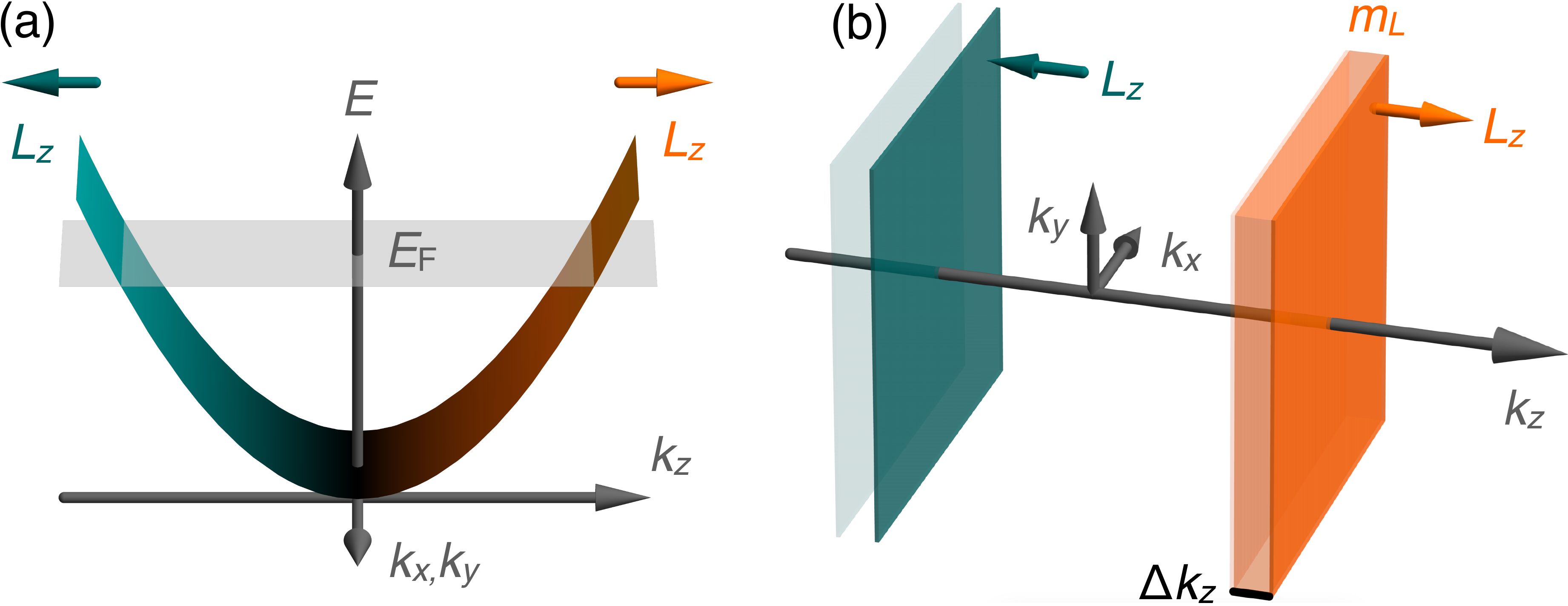}
    \caption{Orbital Edelstein effect (schematic for $c>0$). (a) Band structure of free electrons with orbital angular momentum $L_z$ proportional to the group velocity (color). (b) Upon application of an electric field along $z$, the initial Fermi surfaces (pale) are shifted by $\Delta k_z$ giving rise to a non-equilibrium orbital magnetic moment $\vec{m}_{L}$.}
    \label{fig:edelstein}
\end{figure}

Recently, several phenomena related to angular momentum have been observed and understood based on the orbital degree of freedom. Orbital-polarized currents, generated by the orbital Hall effect~\cite{zhang2005intrinsic, bernevig2005orbitronics, kontani2008giant, tanaka2008intrinsic, kontani2009giant,go2018intrinsic, pezo2022orbital,canonico2020orbital,cysne2022orbital,salemi2022theory,busch2023orbital,choi2023observation,lyalin2023magneto,busch2024ultrafast,gobel2024OHE,gobel2024topological} or the orbital EE~\cite{levitov1985magnetoelectric,yoda2015current,yoda2018orbital, go2017toward,salemi2019orbitally,johansson2021spin,liu2021chirality,kim2023optoelectronic,el2023observation,leiva2023spin,hagiwara2024orbital,johansson2024theory,lee2024orbital} can lead to the generation or transport of orbital angular momentum when currents or electric fields are applied. It has been shown that the orbital EE is sensitive to the chirality of a system as well~\cite{yoda2015current,yoda2018orbital,liu2021chirality,kim2023optoelectronic,hagiwara2024orbital}. We refer to this phenomenon as chirality-induced orbital selectivity (CIOS). \change{When a current flows through a chiral material, an orbital magnetic moment is generated due to a shift of the Fermi surfaces, as visualized in Fig.~\ref{fig:edelstein}. If the chiral material is interfaced with another material, an orbital current may flow. This material may serve as an electrode, which typically has a considerable spin-orbit coupling (SOC), for example Pt. This leads to the conversion of the orbital current into a spin current. The orbital Edelstein effect may then be detected as CISS and not as CIOS, where the orbital contribution is scaled by the orbital-to-spin conversion efficiency of the electrode. This contribution occurs additionally to any genuine spin contribution to CISS and could therefore account for the discrepancy of theoretically and experimentally determined CISS efficiencies. Therefore, understanding the mechanism behind the orbital Edelstein effect in chiral materials is of crucial importance.}

In most theoretical descriptions, orbital angular momentum is calculated via the hybridization of multiple atomic orbitals at the same atom (atomic-center approximation), the efficiency of CIOS is underestimated and can hardly be understood intuitively because the genuine chiral mechanism is not captured: The orbital angular momentum is mostly generated by the circular motion of electrons among \textit{multiple} atoms forming a helix, like in tellurium. This generates an orbital angular momentum that is always along the current direction for one helix chirality and always opposite for the other (cf. arrows in Fig.~\ref{fig:overview}).
\change{Even in a semi-classical analysis, we can understand why considering inter-side hybridization is crucial: Electrons moving along a helix generate a current and orbital angular momentum along the same direction. The orbital-polarized current is independent of the motion direction but changes sign with the chirality of the helix because the angular velocity is proportional to the velocity along the helix and the chirality (cf. Fig.~\ref{fig:overview}). Here it becomes apparent that the orbital angular momentum is generated due to the circular motion among \textit{multiple} atoms forming the helix and is not introduced by hybridization of orbitals of one atom.}
Currently, the literature on the orbital Edelstein effect considering such inter-site contributions is scarce. Refs.~\cite{yoda2015current, yoda2018orbital} use the modern formulation of orbital magnetization ~\cite{chang1996berry,xiao2005berry,thonhauser2005orbital,ceresoli2006orbital,raoux2015orbital,gobel2018magnetoelectric} and even make the analogy to a classical solenoid. However, due to the Haldane-like~\cite{haldane1988model} nature of the model~\cite{yoda2015current, yoda2018orbital}, requiring multiple independent parameters, the Edelstein susceptibility cannot be calculated analytically. Furthermore, the interpretation based on Weyl points in the three-dimensional Brillouin zone dilutes the intuitive mechanism based on the chirality in real space.

In this paper, we introduce the minimal model for understanding \change{the Edelstein effect in chiral materials}. The unit cell of the considered helix consists of only three $s$ orbitals and extends periodically along one dimension. We consider the modern formulation of orbital magnetization for the calculation of the orbital angular momentum to account for inter-site contributions that are crucial for describing the effect in chiral materials properly. Since we can describe this system analytically, we are able to reveal the chiral mechanism of CIOS and CISS based on an orbital EE that is orders of magnitude larger than the spin EE.


\change{This paper is structured as follows. First, we introduce the model and analytically calculate the band structure, orbital angular momentum and the orbital Edelstein susceptibility in Sec.~\ref{sec:model}. Afterwards, in Sec.~\ref{sec:real}, we discuss the consequences of two important effects present in real materials: dispersion perpendicular to the helix direction and spin-orbit coupling. In Sec.~\ref{sec:effectivemodel}, we show that the analytical model can be even further simplified by making use of a constraint between the angular velocity and the velocity along the helix, as in a classical analysis. We discuss our results and conclude in Sec.~\ref{sec:conclusion}.}


\section{Orbital Edelstein effect in analytically solvable model} \label{sec:model}

The model consists of the most fundamental chiral structure: a helix consisting of 3 atoms akin to the structure in tellurium and resembling its space group $P3_121$ ($D_3^4$ symmetry) or $P3_221$ ($D_3^6$ symmetry), depending on the chirality~\cite{furukawa2017observation}. The three basis atoms positioned at $\vec{r}_1=(0,0,0), \vec{r}_2=\left(a,0,\nicefrac{1}{3}\,c\right), \vec{r}_3=\left(\nicefrac{1}{2}\,a,\nicefrac{\sqrt{3}}{2}\,a,\nicefrac{2}{3}\,c\right)$ host a single $s$ orbital and give rise to hopping paths along 
\begin{align}
    \vec{r}_{12}&=\left(a,0,\nicefrac{1}{3}\,c\right),\notag\\
    \vec{r}_{23}&=\left(-\nicefrac{1}{2}\,a,\nicefrac{\sqrt{3}}{2}\,a,\nicefrac{1}{3}\,c\right),\\
    \vec{r}_{31}&=\left(-\nicefrac{1}{2}\,a,-\nicefrac{\sqrt{3}}{2}\,a,\nicefrac{1}{3}\,c\right).\notag
\end{align}
The chiral structure is considered periodic along the $z$ direction with a period of $c$. The sign of $c$ controls the chirality of the helix: right-handed for $c>0$ and left-handed for $c<0$, cf. Fig.~\ref{fig:overview}. We also consider periodicity in the $xy$ plane but for simplicity the helices are not coupled by hoppings. This allows us to describe the chiral structure in an effectively one-dimensional Brillouin zone with 
$k_z$ the only wave vector component that exhibits dispersion. \change{The $x$ and $y$ coordinates of the atoms are important to account for the chirality and for the definition of the orbital angular momentum.}

\subsection{Hamiltonian and band structure}

The tight-binding Hamiltonian requires only a single parameter (nearest-neighbor hopping amplitude $t$) to capture the chiral mechanism of CIOS
\begin{align}
    H(\vec{k})=-t\begin{pmatrix}
        0&
        e^{i\vec{k}\cdot \vec{r}_{12}}&
        e^{-i\vec{k}\cdot\vec{r}_{31}}
        \\
        e^{-i\vec{k}\cdot\vec{r}_{12}}&
        0&
        e^{i\vec{k}\cdot\vec{r}_{23}}
        \\
        e^{i\vec{k}\cdot\vec{r}_{31}}&
        e^{-i\vec{k}\cdot\vec{r}_{23}}&
        0
    \end{pmatrix}.
\end{align}
Due to the lack of dispersion along $k_x$ and $k_y$ the band structure along $k\equiv k_z$ reads
\begin{align*}
    E_1(k)&=-2t\cos\left(k\,\frac{c}{3}\right),\quad\quad E_{2,3}(k)=E_1\left(k\pm\frac{2\pi}{c}\right).
\end{align*}
This band structure [Fig.~\ref{fig:results}(a)] is equivalent to the single band $-2t\cos\left(k\,\frac{c}{3}\right)$ characterizing a one-dimensional chain of atoms with lattice constant $c/3$ along $z$ which has been back-folded into the smaller Brillouin zone accounting for the $3$ atoms in the unit cell. 
\change{Note that the band structure can be very easily calculated by diagonalizing the Hamiltonian for $k_x=k_y=0$.}
The eigenvectors $\ket{n}$ along $k_z$ are constant and read
\begin{align*}
    \ket{1}=\frac{1}{\sqrt{3}}\begin{pmatrix}1\\1\\1\end{pmatrix},\quad\quad
    \ket{2,3}=\frac{1}{\sqrt{3}}\begin{pmatrix}1\\-\nicefrac{1}{2}\pm i\nicefrac{\sqrt{3}}{2}\\-\nicefrac{1}{2}\mp i\nicefrac{\sqrt{3}}{2}\end{pmatrix}.\notag
\end{align*}

\subsection{Orbital angular momentum}

We calculate the $z$ component of the orbital angular momentum of the $n$th band based on the modern formulation of orbital magnetization~\cite{chang1996berry,xiao2005berry,thonhauser2005orbital,ceresoli2006orbital,raoux2015orbital,gobel2018magnetoelectric}
\begin{align}
      L_{n,z}(\vec{k}) &= i \frac{m_e}{g_L\hbar}  \sum_{m \neq n} \frac{1}{E_{m}(\vec{k}) - E_{n}(\vec{k})}\\
      &\times\left[\braopket{n}{\nicefrac{\partial H}{\partial k_x}}{m} \braopket{m}{\nicefrac{\partial H}{\partial k_y}}{n} - (n\leftrightarrow m)\right].\notag
\end{align}
\change{The derivatives of the Hamiltonian $\frac{\partial H}{\partial k_x}$, $\frac{\partial H}{\partial k_y}$ account for the chirality of the system.}
This results in a chirality-dependent orbital angular momentum (opposite for $c>0$ versus $c<0$)
\begin{align}
    L_{1,z}(k)&=\frac{1}{\sqrt{3}}\frac{m_e}{g_L\hbar}a^2 t\,\sin\left(k\frac{c}{3}\right),\\ L_{\nicefrac{2}{3},z}(k)&=L_{1,z}\left(k\pm\frac{2\pi}{c}\right)
\end{align}
that is proportional to the group velocity $v_{n,z}(k)=\frac{1}{\hbar}\frac{\partial E_n}{\partial k_z}$.

\subsection{Orbital Edelstein effect} \label{sec:oee}

\begin{figure}[t!]
    \centering
    \includegraphics[width=1\columnwidth]{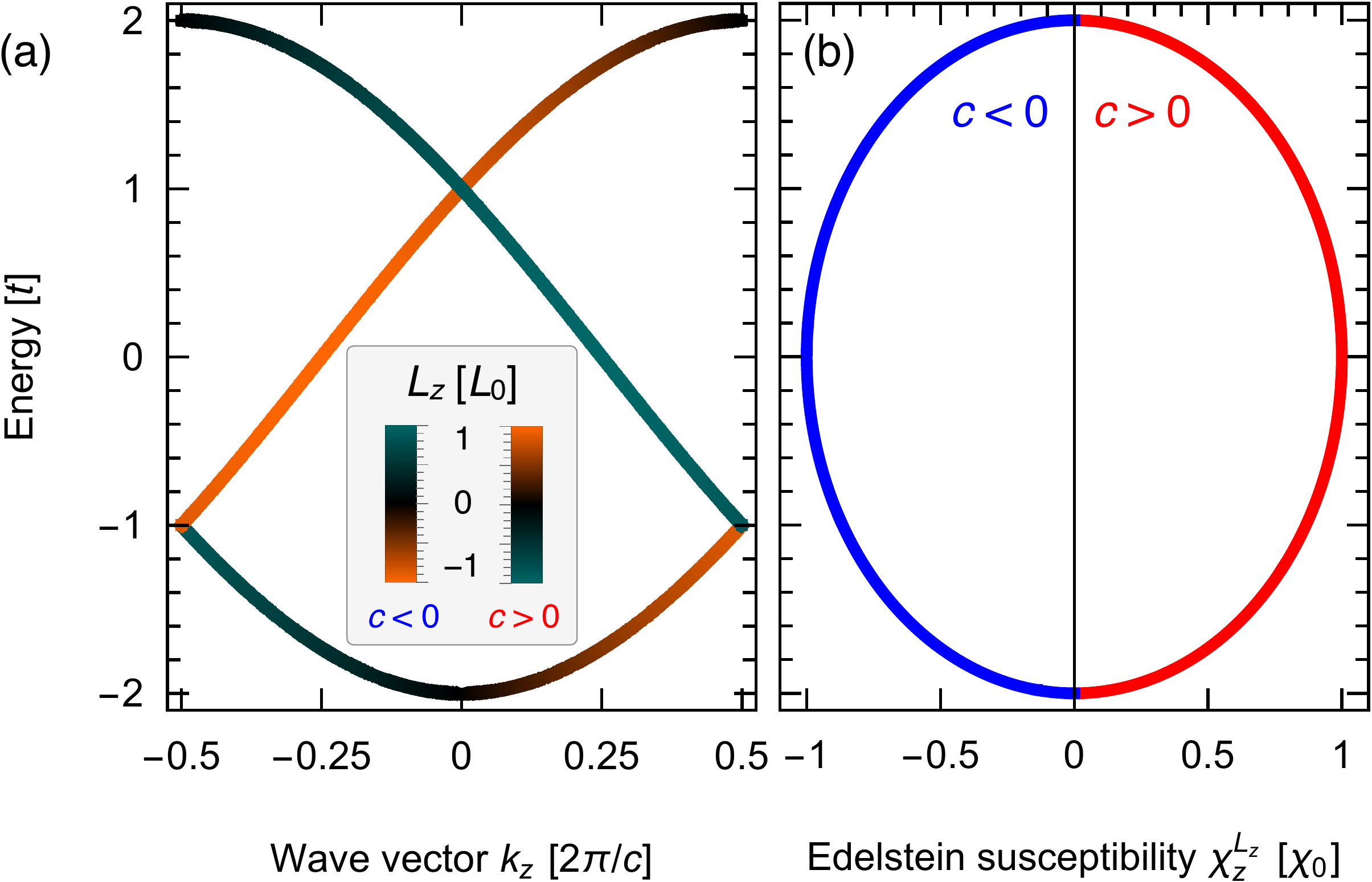}
    \caption{Chirality-induced orbital selectivity in a chiral system without spin-orbit coupling. (a) Band structure for which the color encodes the orbital angular momentum $L_z$ in units of $L_0=\frac{1}{\sqrt{3}}\frac{m_e}{g_L\hbar}a^2 t$. (b) Edelstein susceptibility $\chi_z^{L_z}$ in units of $\chi_0=\frac{e^2}{2\pi \sqrt{3}\,\hbar^2}\tau \,|t|ca^2$. The red curve shows the result for the right-handed system ($c>0$) and the blue curve for the left-handed system ($c<0$).}
    \label{fig:results}
\end{figure}

The fact that the orbital angular momentum is antisymmetric with $k$, while the band structure is symmetric [cf. also Fig.~\ref{fig:results}(a)], gives rise to an orbital Edelstein effect, as visualized in Fig.~\ref{fig:edelstein}.
The collinear orbital Edelstein susceptibility $\chi_z^{L_z}$ relates the applied electric field $\vec{E}=E_z\vec{e}_z$ and the generated non-equilibrium orbital magnetic moment per unit cell $m_{L_z}^\mathrm{uc}$ via
%
$m_{L_z}^\mathrm{uc}=\chi_z^{L_z}E_z$.
%
We can calculate $\chi_z^{L_z}$ easily using the Boltzmann transport theory when we assume a constant relaxation time $\tau$
\begin{align}
    \chi_{z}^{L_z}(E)=\frac{e^2g_L}{2m_e}\tau\sum_{n,\vec{k}} v_{n,z}(k)\cdot L_{n,z}(k)\cdot\delta(E_n(k)-E).
\end{align}
%
%
%
This results in
\begin{align}
    \chi_{z}^{L_z}(E)=\frac{e^2}{2\pi \sqrt{3}\,\hbar^2}\tau \,|t|ca^2\sqrt{1-\left(\frac{E}{2t}\right)^2}.
\end{align}
%
The result is proportional to $L_{n,z}^{(k>0)}(E)-L_{n,z}^{(k<0)}(E)$. Therefore, we can easily understand the energy dependence as presented in Fig.~\ref{fig:results}(b). $\chi_{z}^{L_z}$ is zero at the band edges due to the zero velocity and is largest in the middle of the band width where the velocity and therefore $L_z$ are largest as well. The orbital Edelstein susceptibility changes sign once the chirality is reversed, due to the linear $c$ dependence~\footnote{Note that increasing the helix period $c$ would lead to a decreased hopping amplitude $t$.}.

%
%
%

So far, we wanted to make the discussion as general as possible. Now we consider the case of tellurium's $s$ bands that exhibit an almost perfect resemblance of the band structure we have calculated. We use $a=2.133\,$\AA, $t=1.249\,\mathrm{eV}$~\cite{li2013non} and $g_L=1$ which gives a maximum value for the orbital angular momentum $L_0=0.43\,\hbar$. For the orbital Edelstein susceptibility we additionally use $c=\pm 6.018\,$\AA\, and assume a relaxation time $\tau=1\,\mathrm{ps}$ which results in an extremum of $\chi_0=\pm 125.0\times10^{-9}\,\mu_\mathrm{B}\,\mathrm{m}/\mathrm{V}$. \changeb{Note that the magnitude of this signal scales with the relaxation time which is a parameter in our calculations.} The sign depends on the crystal chirality which means the system exhibits an extremely strong yet purely geometrical \change{orbital Edelstein effect}. In our model $\chi_{z}^{S_z}=0$, because spin is not yet considered.

\section{Additional effects present in real materials} \label{sec:real}

\change{After having established the main result of the paper, we test its validity by comparison with a three-dimensional model (cf. Sec.~\ref{sec:comparison3d}). Afterwards, we present how including SOC influences the orbital Edelstein effect and may cause a spin Edelstein effect as well; two effects relying on distinct mechanisms (cf. Sec.~\ref{sec:soc}).}

\subsection{Hopping between the helices} \label{sec:comparison3d}

We note that we have restricted our model to quasi one dimension, so that we can solve it analytically. In reality, the individual helices are coupled by van der Waals interactions and dispersion along $k_x$ and $k_y$ emerges. However, especially near the center of the energy range, the Fermi surfaces are sheets that extend throughout the whole $k_xk_y$ plane of the three-dimensional Brillouin zone, similar to the schematics shown in Fig.~\ref{fig:edelstein}(b). \change{In Appendix~\ref{app:comparison3d}}, we show that the full three-dimensional system is here described well by the quasi-one-dimensional analytical model.


\subsection{Influence of spin-orbit coupling} \label{sec:soc}
\change{The following discussion is again for the one-dimensional model discussed in Sec.~\ref{sec:model}. To include the effect of SOC, first}, we make the Hamiltonian spin dependent. Without SOC, all bands are now doubly degenerate but the orbital angular momentum always has the same value for the two uncoupled spin-subsystems. Therefore, the orbital Edelstein susceptibility $\chi_{z}^{L_z}(E)$ is simply increased by a factor of 2 to a maximum of $250.0\times10^{-9}\,\mu_\mathrm{B}\,\mathrm{m}/\mathrm{V}$. The spin Edelstein susceptibility $\chi_{z}^{S_z}(E)=0$ due to Kramer's degeneracy.

\begin{figure}[t!]
    \centering
    \includegraphics[width=1\columnwidth]{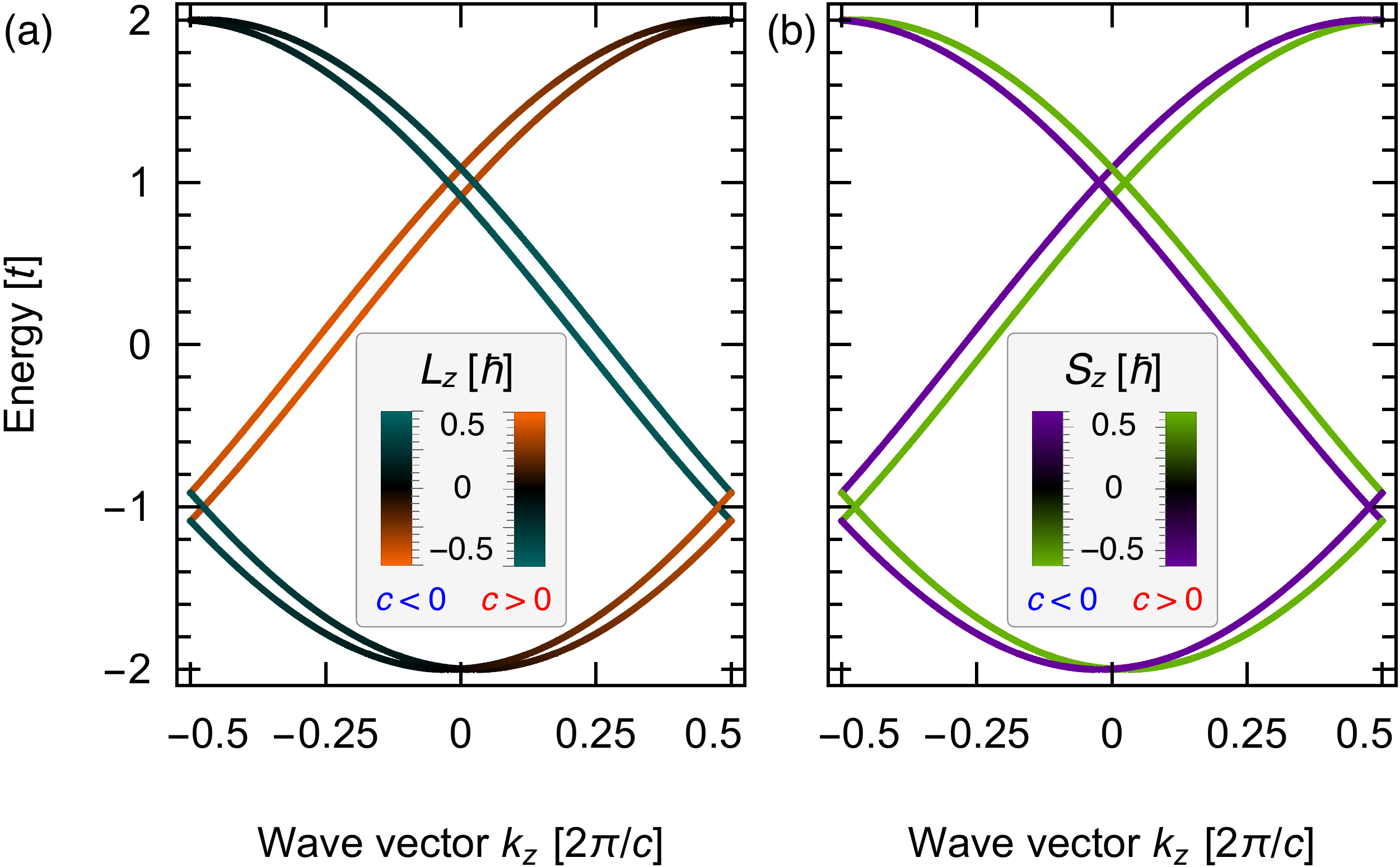}
    \caption{Band structure of a chiral helix with spin-orbit coupling. (a) The color encodes the orbital angular momentum $L_z$. (b) The color encodes the spin $S_z$. Parameters correspond to tellurium with exaggerated spin-orbit coupling $\lambda=0.05t$.}
    \label{fig:resultsSOC}
\end{figure}

\change{SOC is included via a spin- and chirality-dependent hopping term along the helix direction. We introduce the term in Appendix~\ref{app:soc} and explain that it corresponds to a reduced version of the Kane-Mele SOC term~\cite{kane2005z}. The resulting band structure exhibits spin-split bands in which the energy is slightly shifted along the $\pm k_z$ direction. Likewise, the orbital angular momentum is also only slightly affected by this shift along $\pm k_z$ [cf. Fig.~\ref{fig:resultsSOC}(a)]. The analytical calculations are presented in Appendix~\ref{app:soc}.}

\change{For the orbital Edelstein susceptibility, we get the same energy-dependent curve as without SOC but with a rescaled bandwidth $\tilde{t}_\lambda = t\sqrt{1+(\lambda/t)^2}$, where $\lambda$ quantifies the SOC strength; cf. Appendix~\ref{app:soc}.} Even if the SOC were $5\%$ of the hopping amplitude $\lambda=0.05t$, the maximum of the orbital Edelstein susceptibility changes by only $0.125\%$. This highlights that the orbital Edelstein effect is not governed by SOC but purely by the chiral structure of the helices. The contributions of the inner and outer bands to the spin Edelstein susceptibility cancel, $\chi_z^{S_z}=0$. \change{In Appendix~\ref{app:relax}, we analyze} that once a spin EE emerges, e.\,g. by considering a $k$-dependent relaxation time, it is proportional to spin-orbit coupling and orders of magnitude smaller than the orbital EE.


\section{Effective single-band description} \label{sec:effectivemodel}

Lastly, we want to compare the quantum mechanical results with a classical description using constraints. If an electron moves along the helix, it carries out a net motion along $z$ superimposed by a periodic motion along the edges of a triangle in the $xy$ plane. The latter generates an orbital angular momentum $L_z = m_e\frac{a}{2\sqrt{3}}v_{xy}$ with respect to the center of this triangle. The geometry introduces a constraint of the periodic and translational motion: the in-plane velocity along the sides of the triangle $v_{xy}$ is proportional to the velocity along the periodic direction $v_z$, so $v_{xy}=v_z\cdot 3a/c$. This means, the orbital angular momentum and the velocity along $z$ are proportional to each other $L_z^{\mathrm{classical}} = m_e\frac{\sqrt{3}a^2}{2c}v_z$ which is a signature of the system's chirality.

\begin{figure}[t!]
    \centering
    \includegraphics[width=1\columnwidth]{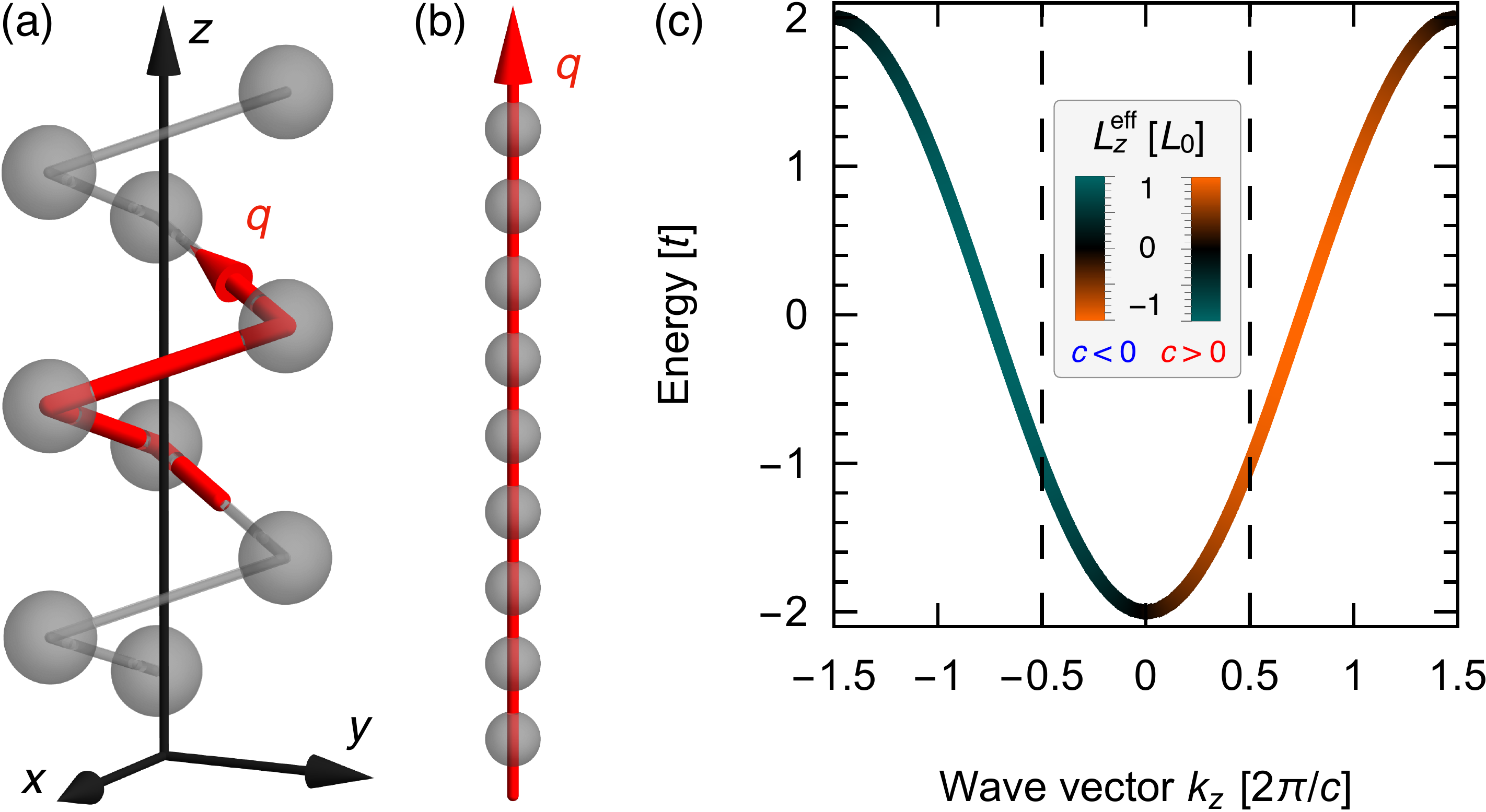}
    \caption{Effective 1-band model. (a) Helix structure for $c>0$ with generalized coordinate $q$ (red). (b) Structure along this new effective coordinate. (c) Band structure with effective orbital angular momentum (color). The dashed line indicates the smaller Brillouin zone corresponding to a chain with 3 atoms in the unit cell.}
    \label{fig:effective}
\end{figure}

If we carry over this result to the quantum mechanical tight-binding model, we can effectively describe the system by a single band (or band-pair when spin is considered). The initial system [Fig.~\ref{fig:effective}(a)] with a generalized coordinate $q$ is mapped to a 1-dimensional chain along $z$ with lattice constant $c/3$ [Fig.~\ref{fig:effective}(b)] and the Brillouin zone is now three times as large along $k_z$ as before. This mapping is possible because the local environment of all three basis atoms in the helix is equivalent: We can go from one atom to another by rotation and translation. 

The 1-dimensional chain by itself is not chiral anymore and motion along the chain would not give rise to an orbital angular momentum. However, we can account for the periodic in-plane motion by introducing an effective orbital angular momentum operator $L_z^{\mathrm{eff}}$. It is derived by replacing the Hamilton function $\mathcal{H}$ with the Hamilton operator $H^{\mathrm{eff}}$ for the calculation of the velocity along the helix direction $v_z=\frac{\partial \mathcal{H}}{\partial p_z}\rightarrow v_z=\frac{1}{\hbar}\frac{\partial H^{\mathrm{eff}}}{\partial k_z}$ leading to
\begin{align}
    L_z^{\mathrm{eff}}=m_e\frac{\sqrt{3}a^2}{2c\hbar}\frac{\partial H^{\mathrm{eff}}}{\partial k_z}.
\end{align}
Together with this operator, the effective Hamiltonian 
\begin{align}
    H^{\mathrm{eff}}(k_z)&=-2t\cos(k_zc/3)\sigma_0- 2\lambda\sin(k_zc/3)\sigma_z
\end{align}
gives rise to the exact same Edelstein susceptibilities $\chi_z^{L_z}(E)$ and $\chi_z^{S_z}(E)$ as the full model considered before, in agreement with the Ehrenfest theorem. This is because $E(k_z)$, $L_z(k_z)$ and $S_z(k_z)$ are the same as before only that the system is now described in a larger Brillouin zone [Fig.~\ref{fig:effective}(c)].
This effective 2-band description (or 1-band description without spin) shows that a semi-classical explanation of CIOS is applicable and allows us to understand even easier why the orbital EE is much larger than the spin EE.

Note, that this analogy is only truly valid when the helices are not coupled by hopping terms so that $L_z\propto v_z$. However, as we have shown in \change{Appendix~\ref{app:comparison3d}}, the analytical results even agree well with the numerical calculations based on a full three-dimensional description.

\section{Conclusion} \label{sec:conclusion}

We have shown that the orbital Edelstein effect in a helix is generated by non-local intersite contributions and is not generated by the hybridization of quantum-mechanical orbitals at a single atom. This new mechanism is fundamentally different from previously proposed mechanisms because it originates purely from the chirality of the system and can be understood even classically without the need for spin-orbit coupling.

We have analytically shown that the current-induced selectivities of spin and orbital angular momentum in chiral materials follow fundamentally different mechanisms. The orbital contributions from bands with opposite spin directions add up and the effect occurs even without SOC. \changeb{Within our model, the spin EE of both bands (almost) cancels and scales with SOC making it a less efficient mechanism. However, we note that additional contributions to the spin EE can emerge when terms are added that we have not considered in our minimal model. For example, the model calculations of Ref.~\cite{calavalle2022gate} consider not only hopping terms in the plane between the helices but also an in-plane spin-momentum locking. To test the applicability of our model and the validity of the conclusions, it will be crucial to distinguish the spin and orbital contributions to the measured Edelstein effect in the future, for example, in tellurium.}

\change{The generated Edelstein effect is also very different from the conventional Rashba-Edelstein effect, typically generated by the Rashba interaction at interfaces, because: (i) the generated current-induced orbital magnetic moment is oriented along the current direction (versus perpendicular for Rashba). (ii) It is a non-local effect that occurs all throughout the helix (versus localized at the interface for Rashba). (iii) It is caused purely by the chirality of the helix and not by spin-orbit coupling (versus by spin-orbit coupling and a broken inversion symmetry for Rashba). (iv) It has its origin in the real-space chirality (versus relying on spin-momentum locking in reciprocal space for Rashba).}

\change{Our findings are significant for intuitively understanding the mechanism behind the orbital Edelstein effect in chiral materials but also motivate more detailed research to find ideal chiral materials that can be used for applications. A large orbital Edelstein effect can be used to generate large torques that are important for spin-orbitronic devices. For example, in Ref.~\cite{gobel2025chirality} we have shown that in chiral carbon nanotubes the mechanism for the orbital Edelstein effect is in principle the same as we have presented here. However, there exist many different chiral hopping paths along a chiral nanotube. Therefore, a direct relation to the semi-classical explanation of the effect is not always possible and the influence of the complicated sub-band structure has to be considered as well for most energies.}

\change{Recently, several options for the detection of orbital magnetic moments and orbital currents have been demonstrated~\cite{jo2024spintronics}: (i) By measuring orbital torques as in Ref.~\cite{lee2021orbital}, (ii) Via magneto-optical Kerr effect measurements as in Ref.~\cite{choi2023observation,lyalin2023magneto}, (iii) By detecting charge currents via inverse orbital effects as in Ref.~\cite{el2023observation}, (iv) Via the Hanle magnetoresistance as in Ref.~\cite{sala2023orbital}. We would like to highlight here especially Ref.~\cite{el2023observation} in which the inverse orbital Edelstein effect has been measured at a LaAlO$_3$/SrTiO$_3$ interface.}
\change{Furthermore, as explained in the introduction, if a spin signal is measured (as for detecting CISS), the current-induced orbital moment will lead to an orbital current that will be converted into a spin current upon passing through the detecting electrode. So, CIOS might effectively be detected as CISS in addition to any genuine CISS signal. Therefore, our work helps filling the gap between the experimentally observed signals and the too small Edelstein signals that were calculated before based only on the spin EE or the orbital EE effect in the atomic-center approximation.}

\section*{Appendices}
\appendix

\section{Comparison with a three-dimensional model} \label{app:comparison3d}

In the main text of this paper, the helices are not coupled by hoppings which leads to flat bands along $k_x$ and $k_y$. 

\begin{figure*}[th!]
    \centering
    \includegraphics[width=\textwidth]{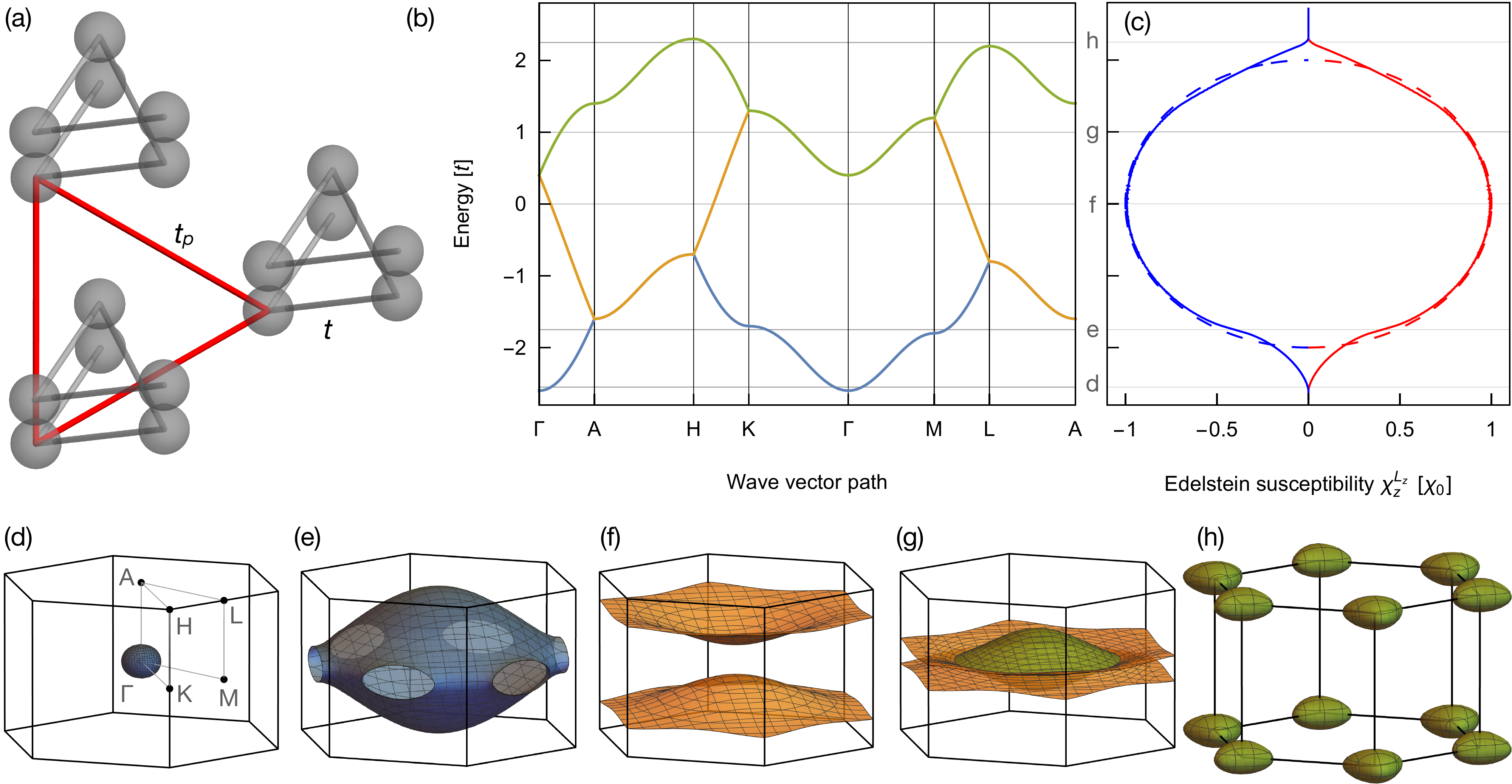}
    \caption{Chirality-induced orbital selectivity in a three-dimensional model. (a) Example of additional hopping paths considered in this calculation (red). (b) Band structure for $t_p=0.1t$. (c) Orbital Edelstein susceptibility in units of $\chi_0=\frac{e^2}{2\pi \sqrt{3}\,\hbar^2}\tau \,|t|ca^2$ that has been calculated numerically for positive (red) and negative (blue) chirality. The dashed line corresponds to the analytical results derived from the one-dimensional model ($t_p=0$), as presented in Fig.~\ref{fig:results}(b). (d-h) Fermi surfaces at energies $-2.55t$, $-1.75t$, $0t$, $1t$, $2.25t$, respectively.}
    \label{fig:SM3d_inplane}
\end{figure*}

\subsection{In-plane hopping between helices}

First, we couple the helices by considering hoppings in the $xy$ plane, as presented in Fig.~\ref{fig:SM3d_inplane}(a). The band structure [Fig.~\ref{fig:SM3d_inplane}(b)] exhibits dispersion along the $k_x$ and $k_y$ directions and the model is not effectively one-dimensional anymore.

Still, the energy-dependent orbital Edelstein susceptibility [Fig.~\ref{fig:SM3d_inplane}(c)] is described by the one-dimensional model (dashed lines) exceptionally well. This is especially the case in the energy range $-1.7t\lesssim E_\mathrm{F}\lesssim 1.4t$ where the Fermi surfaces are nearly planar sheets [Fig.~\ref{fig:SM3d_inplane}(f)] that extend throughout the entire $k_xk_y$ plane as in Fig.~\ref{fig:edelstein}(b). Near the band edges, the Edelstein susceptibility deviates from the analytical result derived from the one-dimensional model because the dispersion along $k_x$ and $k_y$ causes the Fermi surfaces to turn into deformed spheres [Fig.~\ref{fig:SM3d_inplane}(d,h)] whose surface area is finite and changes considerably with energy. This effect is not captured by the one-dimensional model.

\subsection{Chiral second-nearest neighbor hopping}

\begin{figure*}[th!]
    \centering
    \includegraphics[width=\textwidth]{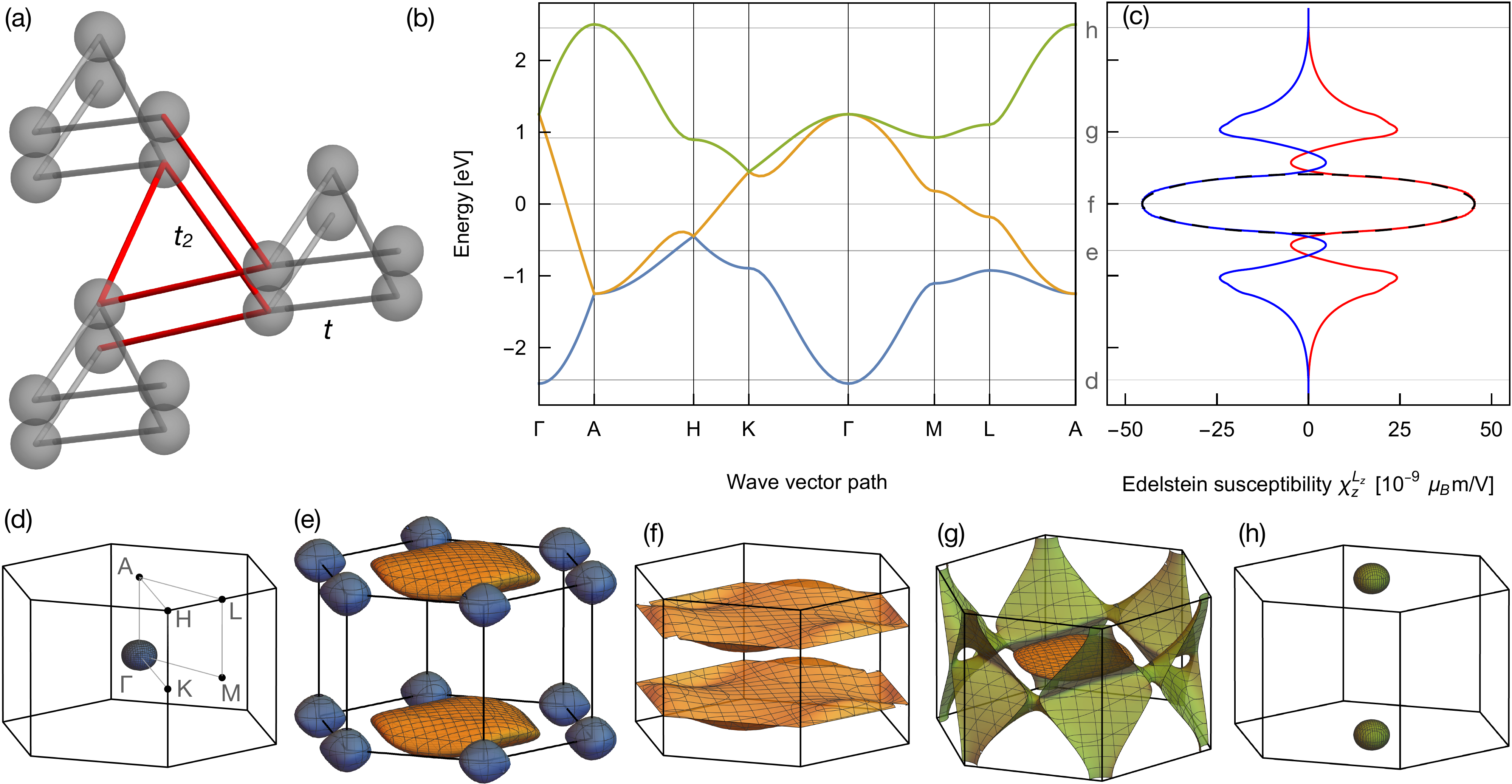}
    \caption{Chirality-induced orbital selectivity in a three-dimensional model. (a) Example of additional hopping paths considered in this calculation (red). (b) Band structure of the $s$ bands of tellurium. (c) Orbital Edelstein susceptibility that has been calculated numerically for positive (red) and negative (blue) chirality. The dashed line shows a fit using the analytically derived result based on the one-dimensional model. (d-h) Fermi surfaces at energies $-2.45\,\mathrm{eV}$, $-0.65\,\mathrm{eV}$, $0\,\mathrm{eV}$, $0.92\,\mathrm{eV}$, $2.45\,\mathrm{eV}$, respectively. The hopping amplitudes are $t=0.71487\,\mathrm{eV}$ and $t_2=0.267566\,\mathrm{eV}$, fitting the band structure of Ref.~\cite{li2013non}.}
    \label{fig:SM3d_tellurium}
\end{figure*}

For the lattice parameters $a=2.133\,$\AA, $b=4.461\,$\AA\, (distance between the helices) and $c=6.018\,$\AA\, of tellurium, the second-nearest neighbor hoppings are not the hoppings in the $xy$ plane but the ones indicated in Fig.~\ref{fig:SM3d_tellurium}(a). This gives rise to a richer energy dependence of the Edelstein susceptibility that exhibits minima and maxima [Fig.~\ref{fig:SM3d_tellurium}(c)] and to Fermi surfaces that consist of multiple different shapes [Fig.~\ref{fig:SM3d_tellurium}(e,g)].

Still, the Edelstein susceptibility can be fitted by the results derived from the analytical model in the main text [dashed lines in Fig.~\ref{fig:SM3d_tellurium}(c)]. This works especially well near the center of the energy range where the Fermi surfaces are still nearly planar sheets [Fig.~\ref{fig:SM3d_tellurium}(f)] that extend throughout the whole $k_xk_y$ plane as in Fig.~\ref{fig:edelstein}(b).

The newly considered hoppings with amplitude $t_2$ are chiral and couple the dispersion along $k_z$ with the dispersion along the $k_x$ and $k_y$ directions. Due to the interplay with the other chiral hoppings with amplitude $t$, that have been considered before, the velocity along $z$ and the orbital angular momentum $L_z$ are not always proportional to each other and can even have opposite signs. This gives rise to the minima and maxima in the Edelstein susceptibility [Fig.~\ref{fig:SM3d_tellurium}(c)] that can be fitted with distinct parameters, respectively.

\section{Influence of spin-orbit coupling} \label{app:soc}

\change{SOC is included akin to a reduced Kane-Mele term~\cite{kane2005z} by using a spin-dependent hopping along the helix. The term} can be written in second quantization
\begin{align}
    i\lambda\sum_{\braket{i,j}}\nu_{ij}c_i^\dagger\sigma_z c_j.\label{eq:KaneMele}
\end{align}
Here, $c_i^\dagger$ and $c_j$ are the creation and annihilation operators at sites $i,j$. $\nu_{ij}=\pm 1$ accounts for the hopping path along the helix: $+1$ for a counter-clockwise rotation and $-1$ for a clockwise rotation, so $+\mathrm{sign}(c)$ for hopping along $+z$ and $-\mathrm{sign}(c)$ for hopping along $-z$. 
\change{In our $6\times 6$ matrix this leads to a rescaling of the hopping: $t\rightarrow t_{\pm,c}=(t\pm i\lambda)$ where the sign is determined by the spin orientation and the chirality of the helix $\mathrm{sign}(c)$.}

The resulting band structure [Figs.~\ref{fig:resultsSOC}(a,b)] is
\begin{align}
    E_{1}^{\uparrow\downarrow}(k)&=-2t\cos\left(k\frac{c}{3}\right)\mp 2\lambda \sin\left(k\frac{c}{3}\right),\notag\\
    E_{\nicefrac{2}{3}}^{\uparrow\downarrow}(k)&=E_{1}^{\uparrow\downarrow}\left(k\pm\frac{2\pi}{c}\right).
\end{align}
\change{SOC leads to a shift of the bands along the $\pm k_z$ direction}. Note the sign change of the SOC term for $c>0$ and $c<0$: The sine function comes from adding the two hopping terms along $\pm c\vec{e}_z$ where the upwards hopping path along the helix is rotating oppositely to the downwards hopping path: $i[\exp(ick_z)-\exp(-ick_z)]$. Furthermore, it depends on the spin-orientation. Close to the $\Gamma$ point the SOC-induced term is $\mp2\lambda k_zc/3$, a term that has been derived in Ref.~\cite{yu2020chirality} as well.

The Hamiltonian is separated into spin-up and spin-down blocks meaning that all bands are $\pm 100\,\%$ spin polarized
$S_{n,z}^{\uparrow\downarrow}(k)=\pm\frac{\hbar}{2}$ [color in Fig.~\ref{fig:resultsSOC}(b)].
This makes the system under consideration similar to a Dresselhaus system~\cite{dresselhaus1955spin}, where the structural chirality breaks the inversion symmetry, but in one dimension. The spin is (anti-) parallel to $\vec{k}$; a feature well known from the Fermi surface of tellurium so the expected spin Edelstein effect is collinear~\cite{furukawa2017observation,furukawa2021current,calavalle2022gate,roy2022long,slawinska2023spin} meaning that electric field and generated magnetic moment are collinear. 

In contrast, the orbital angular momenta of the two bands of each spin-split band pair are almost the same [Fig.~\ref{fig:resultsSOC}(a)]. They are affected in a similar way as the band structure 
\begin{align}
    L_{1,z}^{\uparrow\downarrow}(k)&=\frac{1}{\sqrt{3}}\frac{m_e}{g_L\hbar}a^2\left[t\sin\left(k\frac{c}{3}\right)\mp\lambda \cos\left(k\frac{c}{3}\right)\right],\notag\\ 
    L_{\nicefrac{2}{3},z}^{\uparrow\downarrow}(k)&=L_{1,z}^{\uparrow\downarrow}\left(k\pm\frac{2\pi}{c}\right)
\end{align}
and so the orbital Edelstein susceptibility is merely rescaled due to the new bandwidth 
$\tilde{t}_\lambda = t\sqrt{1+(\lambda/t)^2}$
\begin{align}
    \chi_{z}^{L_z}(E)=\frac{e^2}{\pi \sqrt{3}\,\hbar^2}\tau ca^2 \,|\tilde{t}_\lambda|\sqrt{1-\left(\frac{E}{2\tilde{t}_\lambda}\right)^2}.
\end{align}

\change{Note that the SOC term [Eq.~\eqref{eq:KaneMele}] contains only $\sigma_z$ while a strict SOC term according to the Kane-Mele model~\cite{kane2005z} would include also $\sigma_x$ and $\sigma_y$. The preferred spin orientation is not anymore perfectly aligned with the helix direction but slightly tilted along the hopping path for each nearest-neighbor hopping. However, since each Bloch state is homogeneously distributed over all 3 atoms in the unit cell in our model, this tilting cancels after three nearest-neighbor hoppings within a unit cell. The spin expectation value does not have an in-plane component but is slightly reduced from $\pm \hbar/2$. We disregard this small quantitative effect by neglecting the $\sigma_x$ and $\sigma_y$ terms, similar to what has been done in Ref.~\cite{yu2020chirality} because then we are still able to solve the system analytically and because it does not qualitatively affect our results.}

\begin{figure*}[th!]
    \centering
    \includegraphics[width=0.78\textwidth]{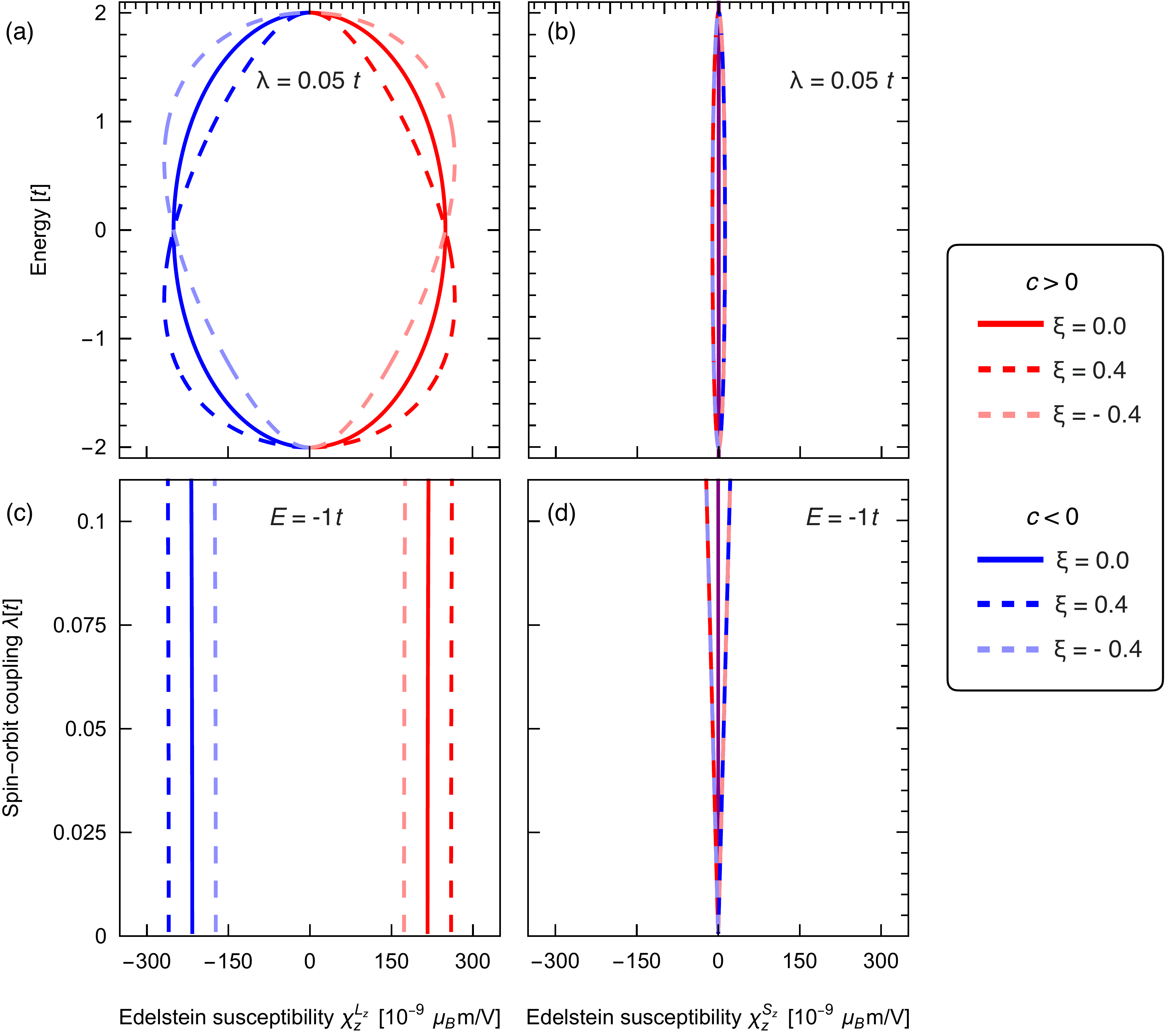}
    \caption{Parameter scaling of the Edelstein susceptibilities. (a) Orbital Edelstein susceptibility $\chi_z^{L_z}$ and (b) spin Edelstein susceptibility $\chi_z^{S_z}$ for a fixed spin-orbit coupling $\lambda=0.05t$ and different values for the $k$-dependent relaxation time $\xi$ (see legend). The energy-dependent curves are opposite for positive (red) and negative (blue) chirality. (c,d) Edelstein susceptibilities as a function of spin-orbit coupling $\lambda$ for a fixed location of the Fermi energy $E=-1t$. Parameters are chosen as in Fig.~\ref{fig:resultsSOC} and correspond to tellurium.}
    \label{fig:SMresultsSOC}
\end{figure*}

\section{Edelstein effect under $k$-dependent relaxation time} \label{app:relax}

To distinguish the here predicted CIOS from the genuine CISS, we calculate the spin Edelstein susceptibility
\begin{align}
    \chi_{z}^{S_z}(E)=\frac{e^2g_S}{2m_e}\tau\sum_{n,\vec{k}} v_{n,z}(k)\cdot S_{n,z}(k)\cdot\delta(E_n(k)-E).
\end{align}
In the considered constant relaxation time approximation, SOC does not lead to a spin Edelstein effect because the contributions of the spin up and down bands cancel\change{, as has been explained in the main text and in Appendix~\ref{app:soc}.}

To generate such an effect, we have several options but all rely on an interplay of multiple effects and therefore generate a rather small signal compared to the orbital Edelstein effect. An example is to consider a conventional Rashba term in the $xy$ plane additionally to the chiral SOC term along $z$, similar to what has been done in Ref.~\cite{calavalle2022gate}. Here we want to present another possibility and consider a $k$ dependent relaxation time. 

Going one order in $k$ beyond the constant relaxation time and accounting for periodicity, it is $\tau_k=\tau_0\left[1+\xi \cos\left(k\frac{c}{3}\right)\right]$. The resulting Edelstein susceptibilities are 
\begin{align}
    \chi_{z}^{S_z}(E)&=-\frac{e^2g_S}{2m\pi}\tau\xi c \,\frac{\lambda}{|\tilde{t}_\lambda|}\sqrt{1-\left(\frac{E}{2\tilde{t}_\lambda}\right)^2},\\
    \chi_{z}^{L_z}(E)&=\frac{e^2}{\pi \sqrt{3}\,\hbar^2}\tau ca^2 \,\left(|\tilde{t}_\lambda|-\frac{1}{2}\frac{t}{|\tilde{t}_\lambda|}E\xi\right)\sqrt{1-\left(\frac{E}{2\tilde{t}_\lambda}\right)^2}.\notag
\end{align}
Note that now there is a finite $\chi_{z}^{S_z}$ and its sign depends on the chirality, the sign of $\xi$ and the sign of the SOC. Still, the orbital contribution dominates and is only slightly affected by SOC. These quantities and their parameter scaling are shown in Fig.~\ref{fig:SMresultsSOC}.

\section*{Acknowledgements}
This work was supported by the EIC Pathfinder OPEN grant 101129641 ``Orbital Engineering for Innovative Electronics''. B.G. and L.S. performed calculations.
B.G. did the analytical derivations, prepared the figures and planned and supervised the project.
All authors discussed the results.
B.G. wrote the manuscript with significant inputs from all authors.
I.M. provided the funding.

\bibliography{short,MyLibrary}

\end{document}